\documentstyle[12pt]{article}
\begin{document}
\title{Universe of Fluctuations II}
\author{B.G. Sidharth$^*$\\
Centre for Applicable Mathematics \& Computer Sciences\\
B.M. Birla Science Centre, Adarsh Nagar, Hyderabad - 500 063 (India)}
\date{}
\maketitle
\footnotetext{$^*$Email:birlasc@hd1.vsnl.net.in; birlard@ap.nic.in}
\begin{abstract}
We throw further light on a recently discussed Kerr-Newman formulation of
Fermions and a related cosmological scheme which predicted an ever
expanding universe, as indeed has subsequently been confirmed. In the
spirit of the correspondence principle, it is shown how the quark
picture emerges at the Compton wavelength and the Big Bang scenario at the
Planck length. At the sametime we obtain a theoretical justification for
the peculiar characteristics of the quarks namely their fractional charge,
handedness and confinement, as also the order of magnitude of their masses,
all of which were hitherto adhoc features.
\end{abstract}
\section{Introduction}
In a previous communication\cite{r1} we considered a Kerr-Newman
type formulation of elementary particles and at the same time a cosmological
scheme that not only deduced from theory, the so called large number
coincidences, but also the age, mass and radius of the universe, the
Hubble constant as also a hitherto adhoc relation between
the pion mass and the Hubble constant which has been described as mysterious by
Weinberg\cite{r2} in a scheme consistent with observation. The model
also predicted an ever expanding universe, as indeed,
subsequent independent supernovae observations have confirmed\cite{r3,r4}.\\
The cosmological scheme was based on the concept of space time quantization\cite{r5,r6},
at the root of which is, the negation of arbitrarily
small space time intervals, as indeed Heisenberg's Uncertainity Principle
demands.\\
It is possible to come to this conclusion by the treatment of elementary
particles, for example electrons, as Kerr-Newman type Black Holes bounded
by the Compton wavelength\cite{r7}. Indeed the remarkable fact has
been well known that the Kerr-Newman metric describes the field of the
electron including the anomalous gyro magnetic ratio $g = 2$\cite{r8}
But there is in this case a naked singularity-- the fact that the radius
of the horizon becomes complex. However it has been shown that in the case
of the electron, this poses no real problem, because in any case, an electron
is physically meaningful only after an averaging of the Zitterbewegung effects
over the Compton wavelength\cite{r9}: In other words the apparent difficulty
arises only if we persist with the purely classical concept of space time
points, and it disappears when these points are replaced by the minimum space
time intervals, which are the physically meaningful quantities. (This point
will be commented upon a little later).\\
In the light of the above, and in the spirit of the correspondence principle
we will now see that the following unified scheme
emerges:\\
i) At the Compton wavelength scales, we can recover the quark picture with a
theoretical basis for hitherto empirical features.\\
ii) At the Planck length, we can recover from the above cosmological scheme the standard
Big Bang scenario.
\section{Quarks}
The puzzling and characteristic feature of quarks are their fractional charge,
handedness and confinement. As noted by Salam and others\cite{r10}, there is no
theoretical basis for these adhoc features. However, exactly such a theoretical
rationale can be obtained, as we will now see.\\
It has been shown that the Kerr-Newman metric for the electron can be
obtained from the linearized theory of General Relativity\cite{r1,r7,r11}.
The starting point is the well known relation\cite{r8}
\begin{equation}
g_{\mu v} = \eta_{\mu v} + h_{\mu v}, h_{\mu v} = \int \frac{4T_{\mu v}
(t - |\vec x - \vec x'|,\vec x')}{|\vec x - \vec x'|} d^3 x'\label{e1}
\end{equation}
with the usual notation.\\
It has been shown in the above references that starting from (\ref{e1}) and
treating the Compton wavelength as the boundary, we can recover for greater
distances, apart from the Quantum Mechanical spin half, the electrostatic potential
viz.,
\begin{equation}
\frac{ee'}{r} = A_0 \approx \frac{2\hbar}{r}\int \eta^{\mu \nu}
\frac{d}{d\tau}T_{\mu \nu} d^3 x'\label{e2}
\end{equation}
where $e'$ is the test charge and also the well known ratio of the electromagnetic
and gravitational interactions, viz.,
\begin{equation}
e^2/m^2 \sim 10^{40}\label{e3}
\end{equation}
However it was also shown that at or near the Compton wavelength itself
equation (\ref{e1}) goes over to
\begin{eqnarray}
4 \quad \eta^{\mu v} \int \frac{T_{\mu \nu} (t,\vec x')}{|\vec x - \vec x' |} d^3 x' +
(\mbox terms \quad independent \quad of \quad \vec x), \nonumber \\
+ 2 \quad \eta^{\mu v} \int \frac{d^2}{dt^2} T_{\mu \nu} (t,\vec x')\cdot |\vec x - \vec x' |
d^3 x' + 0 (| \vec x - \vec x' |^2) \propto - \frac{\propto}{r} + \beta r\label{e4}
\end{eqnarray}
Equation (\ref{e4}) resembles the standard QCD potential\cite{r12}, and this
suggests that the quark model would follow at this scale. The question is,
firstly how do the characteristic and puzzling fractional quark charges emerge,
and secondly, how do their masses turn up?\\
We first observe that the three dimensionality of space, as is well known,
arises from the double connectivity or spin half property\cite{r13,r14}. This spinorial property,
on the other hand is essentially Quantum Mechanical and as has been
discussed\cite{r11}, arises due to the fact that within the Compton wavelength
of the electron, the negative energy components of the Dirac four spinor
dominate, while outside, it is the positive energy components that dominate.
This can also be seen above in the Kerr-Newman type Black Hole description
of the electron.\\
However as we approach the Compton wavelength, we encounter
only negative energy components, and the spinorial and hence three
dimensional feature disappears\cite{r15,r16}. In fact we are in the low
dimensional case. As a confirmation, it is indeed remarkable that such low
dimensional Quantum Mechanical behaviour has already been observed with nano
tubes\cite{r17,r18}.\\
We now consider the electrostatic potential given in (\ref{e2}), but this time
in two dimensions or one dimension. We also use the well known fact that\cite{r19}
in (\ref{e2}) or (\ref{e4}), each of the $T_{\imath j}$ is
given by $\frac{1}{3} \epsilon$, where $\epsilon$ is the energy density. It
now follows from (\ref{e2}) that in two dimensions the particle would have
the charge $\frac{2}{3} e$, or in one dimension it would have the charge
$\frac{1}{3} e$, exactly as in the case of quarks. It also follows that
these fractionally charged particles cannot be observed because by their
very nature they are confined, as indeed is the case. This is also expressed
by the confining part of the QCD type potential (\ref{e4}).\\
To get an idea of the masses of these fractionally charged particles, we
proceed as follows. We consider the QCD type potential (\ref{e4}), but to
facilitate comparison with the usual literature\cite{r20}, we take
$c = \hbar = 1$, and also introduce the factor $\frac{1}{m}$
(owing to the factor $\frac{\hbar^2}{2m}$). The potential (\ref{e4}) now
becomes
\begin{equation}
\frac{4}{m} \quad \eta^{\mu v} \int \frac{T_{\mu v}}{r} d^3 x + 2m \quad \eta^{\mu v}
\int T_{\mu v} r d^3x \equiv -\frac{\propto}{r} + \beta r\label{e5}
\end{equation}
Using the well known relation\cite{r8},
$$m = \int T^{00}d^3x$$
we can see from (\ref{e5}) that $\alpha \sim 0(1)$ and $\beta \sim 0 (m^2)$, where
$m$ is the mass of the quark. This is indeed the case for the QCD
potential. Furthermore it follows from (\ref{e5}) that at the Compton
wavelength scale $r \sim \frac{1}{m}$ (in natural units), the energy is
$\sim 0(m)$. All this is perfectly consistent.\\
We now observe that it follows from (\ref{e2}) (cf.also\cite{r7}), that
\begin{equation}
\frac{e^2}{r} = 2m_e \int \eta^{\mu v}\frac{T_{\mu v}}{r}d^3x\label{e6}
\end{equation}
where $m_e$ is the electron mass. However at the scale of quarks, $e^2$ is
replaced by $\frac{e^2}{10} = \frac{1}{1370} \sim 10^{-3}$. So (\ref{e6})
becomes,
$$\frac{10^{-3}}{r} = 2m_e \int \eta^{\mu v} \frac{T_{\mu v}}{r} d^3 x$$
or
\begin{equation}
\frac{\propto}{r} \sim \frac{1}{r} \approx 2.10^3 m_e \int \eta^{\mu v}
\frac{T_{\mu v}}{r} d^3x\label{e7}
\end{equation}
Comparing (\ref{e7}) with the potential (\ref{e6}),
it follows that the fractionally charged particle which
we have identified with the quark has a mass of the order of $10^3 m_e$ which is correct.\\
We next deduce another interesting property of the quarks from our model.
As noted earlier in this case we are at the Compton wavelength scale, where
we encounter predominantly the negative energy components of the Dirac four spinor, which,
as is well known, have the opposite parity\cite{r21}. So the
quarks should display neutrino like helicity, which indeed is known to be
true, and in the usual theory is attributed to the small Cabibo angle.\\
Thus one could deduce from the above Kerr-Newman type Black Hole model, at
the Compton wavelength scale the fractional charges of the quarks, their confinement,
masses and also their handedness and the QCD potential, all of which were
hitherto adhoc features.
\section{The Big Bang Scenario}
We first recapitulate the main features of the cosmological model described
in the references\cite{r1,r5}, already alluded to. In this model $N$,
the number of elementary particles, typically
pions, in the universe was the only large scale cosmological parameter while
the well known microphysical constants\cite{r22} viz., $\hbar, c,
e$ and the mass (or equivalently the Compton wavelength) of the pion
are taken for granted.\\
It was proposed that in the spirit of Prigogine's cosmology\cite{r23,r24}
from a background Zero Point Field, particles, typically
pions are fluctuationally created within space time intervals ($l, \tau)$,
the Compton wavelength and Compton time of the pion. The rationale for this
is the fact that the energy density of the ZPF fluctuations are\cite{r8},
over a region of dimensions $\lambda$ given by
$\frac{\hbar c}{\lambda^4}$. Whence if $\lambda \sim$ the Compton wavelength
$l$, then the energy of the created particle $\sim mc^2$, where $m$ is the
pion mass.\\
From here it follows that the mass of the universe is given by
$$M = Nm,$$
which indeed is so, remembering that $N \sim 10^{80}$. Further we used the
well known relation\cite{r25},
$$R = \frac{GM}{c^2},$$
where $R$ is the radius of the universe. This again gives the correct value.\\
We next use the fact that given $N$ particles at any epoch, $\sqrt{N}$ particles
are fluctuationally created (cf. also \cite{r25}). Thus, as $\tau$ is the minimum
time unit,
\begin{equation}
\frac{dN}{dt} = \frac{\sqrt{N}}{\tau} = \frac{mc^2}{\hbar} \sqrt{N}\label{e8}
\end{equation}
so that on integration from $t = 0, N = 0$ to the present epoch we get
$$\sqrt{N} = \frac{2mc^2}{\hbar} T$$
which also gives the correct age of the universe viz., $10^{17}secs$.\\
It was also shown that the Hubble constant is given by
$$H = \frac{Gm^3c}{\hbar^2}$$
which is also correct (cf.references\cite{r1}and\cite{r5}). Equivalently
we have
\begin{equation}
m = (\frac{\hbar^2 H}{Gc})^{1/3}\label{e9}
\end{equation}
What makes equation (\ref{e9}) remarkable, is the inexplicable fact that
it relates the pion mass to the Hubble constant, as noted by Weinberg. It
was known purely as an empirical result, but is seen to arise from the present
theory quite naturally.\\
The cosmological constant is given quite consitently
by the relation
$$\wedge \le H^2$$
The model thus predicts an ever expanding and possibly accelerating universe. As
pointed out in the introduction this has since been confirmed by observations
by independent teams. Indeed, apart from this, quite recently the available
data was construed to imply exactly such an expansion\cite{r26}.\\
Moreover in the above model the supposedly mysterious large number coincidences
which lead to Dirac's large number hypothesis are actually consequences of
the theory.\\
Proceeding further we observe that, as noted in the introduction the
concept of space time points is a classical approximation and cannot be reconciled
with Quantum Mechanics even though the latter has unhappily coexisted with this
classical concept. On the other hand the spin half Fermion is a typically
Quantum Mechanical concept, which has not been introduced into classical
General Relativity, as noted by Wheeler and others (cf. references). Infact
as noted in the references\cite{r1,r11}, a reconciliation
is possible only if we consider a minimum time unit, the chronon or the
Compton time and similarly also the minimum length, the Compton wavelength.\\
We recover our classical concept of space time when this minimum interval
tends to $0$. We can immediately see that we then get the Big Bang scenario: In
equation (\ref{e8}), as $\tau \to 0, \frac{dN}{dt} \to \infty$, corresponding
to a singular and instantaneous creation of particles.\\
More accurately we will now consider the limit, not to $0$, but rather to the
Planck scale. The Planck scale corresponds to the extreme classical limit of
Quantum Mechanics, as can be immediately seen from the fact that the Planck
mass $m_P \sim 10^{-5}gms$ corresponds to a Schwarzchild Black Hole of
radius $l_P \sim 10^{-33}cms$, the Planck length. At this stage the spinorial
Quantum Mechanical feature as brought out by the Kerr-Newman type Black
Hole disappears. Infact at the Planck scale we have
\begin{equation}
\frac{Gm_P}{c^2} = \hbar /m_Pc\label{e10}
\end{equation}
In (\ref{e10}), the left side gives the Schwarzchild radius while the right
side gives the Quantum Mechanical Compton wavelength. Another way of writing
(\ref{e10}) is,
\begin{equation}
\frac{Gm^2_P}{e^2} \approx 1,\label{e11}
\end{equation}
Equation (\ref{e11}) expresses the well known fact that the entire energy is gravitational,
rather than electromagnetic, in contrast to equation (\ref{e3}) for a typical
elementary particle.\\
Interestingly from the background ZPF, Planck particles can be produced at the
Planck scales given by (\ref{e10}), exactly as in the case of pions, as seen
earlier. They have been considered to be what may be called a Zero Point
Scale\cite{r27,r28,r29}. But these Planck
particles are much too short lived, with life times $\sim 10^{-42}secs$ and can
at best describe a space time foam\cite{r13}.\\
We will now throw further light on the fact that at the Planck scale it is
gravitation alone that manifests itself. Indeed Rosen\cite{r30} has pointed
out that one could use a Schrodinger equation with a gravitational interaction
to deduce a mini universe, namely the Planck particle. The Schrodinger
equation for a self gravitating particle has also been considered\cite{r31},
from a different point of view. We merely quote the main
results.\\
The energy of such a particle is given by
\begin{equation}
\frac{Gm^2}{L} \sim \frac{2m^5G^2}{\hbar^2}\label{e12}
\end{equation}
where
\begin{equation}
L = \frac{\hbar^2}{2m^3G}\label{e13}
\end{equation}
(\ref{e12}) and (\ref{e13}) bring out the characteristic of the Planck
particles and also the difference with elementary particles, as we will
now see.\\
We first observe that for a Planck mass, (\ref{e12}) gives, self consistently,
$$\mbox{Energy} \quad = m_P c^2,$$
while (\ref{e13}) gives,
$$L = 10^{-33}cms,$$
as required.\\
However, the situation for pions is different: They are parts of the universe
and do not constitute a mini universe. Indeed, if, as above there are $N$ pions
in the universe, then the total gravitational energy is given by, from
(\ref{e12}),
$$\frac{NGm^2}{L}$$
As this equals $mc^2$, we get back as can easily be verified, equation (\ref{e3}),
or equivalently we deduce the pion mass!\\
Indeed given the pion mass, one can verify from (\ref{e13}) that $L = 10^{28}cms$
which is the radius of the universe, $R$. Remembering that $R \approx \frac{c}{H}$,
(\ref{e13}) infact gives back the Weinberg formula, (\ref{e9})!\\
This provides a justification for taking a pion as a typical particle of the
universe, and not a Planck particle, besides re-emphasizing the basic unified
picture of gravitation and electromagnetism.\\
To proceed, let us now use the fact that our minimum space time intervals are $(l_P, \tau_P)$,
the Planck scale, instead of $(l, \tau)$ of the pion, as above.\\
With this new limit, it can be easily verified that the total mass in the volume
$\sim l^3$ is given by
\begin{equation}
\rho_P \times l^3 = M\label{e14}
\end{equation}
where $\rho_P$ is the Planck density and $M$ as before is the mass of the
universe.\\
Moreover the number of Planck masses in the above volume $\sim l^3$ can
easily be seen to be $N' \sim 10^{60}$. However, it must be remembered that
in the physical time period $\tau$, there are $10^{20}$ (that is $\frac{\tau}
{\tau_P})$ Planck life times. In other words the number of Planck particles
in the physical interval $(l, \tau)$ is $N \sim 10^{80}$, the total particle
number.\\
That is from the typical physical interval $(l, \tau)$ we recover the entire
mass and also the entire number of particles in the universe, as in the Big
Bang theory. This also provides the explanation for the above puzzling
relations like (\ref{e14}).\\
That is the Big Bang theory is a characterization of the new model in the
classical limit at Planck scales, but then, we cannot deduce from theory the
relations like the Dirac coincidences or the Weinberg formula.
\section{The Holistic Universe}
In the preceding considerations we had come across the fluctuation in particle
number $\sim \sqrt{N}$ on the one hand and the Quantum Mechanical Uncertainity
length $l$, the Compton wavelength of the pion on the other hand. We can
easily see that these two are not independent, by the following argument\cite{r1,r25}.
The above fluctuation creates an excess
electrostatic potential energy of the electrons viz., $\frac{e^2 \sqrt{N}}{R}$.
When this is balanced by the electron energy, we get
$$\frac{e^2\sqrt{N}}{R} = m_e c^2,$$
which leads to the well known relation,
\begin{equation}
l \approx \frac{R}{\sqrt{N}},\label{e15}
\end{equation}
Thus given $\sqrt{N}$ and $R$, we can deduce $l$. We can appreciate (\ref{e15}),
if we realise that for an assembly of $N$ particles in a volume of radius
$R$, the typical Brownian uncertainity length $l$ is given by (\ref{e15})\cite{r32,r33}.
In other words the Quantum Mechanical Uncertainity arises from the fluctuation
in the number of particles in the universe, in a holistic scenario.\\
Another way of looking at this is, by the following argument.
The
fluctuation in the mass of a typical elementary particle like the pion due to
the fluctuation of the particle number is given by
$$\frac{G\sqrt{N} m^2}{c^2R}$$
So we have
\begin{equation}
(\Delta mc^2) T = \frac{G\sqrt{N}m^2}{R} T = \frac{G\sqrt{N}m^2}{c}\label{e16}
\end{equation}
as $cT = R$. It can be easily seen that the right side of (\ref{e16}) equals $\hbar$!
That is we have
\begin{equation}
\hbar \approx \frac{G\sqrt{N}m^2}{c},\label{e17}
\end{equation}
The equation (\ref{e17}) expresses the Planck constant in terms of non
Quantum Mechanical quantities. This does not mean that $\hbar$ depends
on $N$ or $T$. Rather (\ref{e17}) is another form of equation (\ref{e3}),
as discussed in the references cited. Further the equation (\ref{e16}) is the
well known Quantum Mechanical Uncertainity equation,
$$\Delta E \Delta t \approx \hbar.$$
As mentioned, the equation (\ref{e17}) is
equivalent to the equation (\ref{e3}) expressing the electromagnetism-Gravitational
interaction ratio. In the case of the Planck particle, we had instead of
(\ref{e3}) or equivalently (\ref{e17}) the equation (\ref{e11}) which
expresses the fact that the entire energy is gravitational. Thus
electromagnetism and elementary particles are a result of the fluctuation
of the number of particles in the universe, which leads to Quantum
Mechanical effects. In this case, as seen in Section 3 also, we have
elementary particles being described by the Kerr-Newman metric giving
the full electromagnetic field, instead of the Schwarzchild metric as in
the case of Planck particles.\\
In this light, it is easy to understand equations like (\ref{e9}) which can
relate the pion mass to the Hubble constant.\\
Finally we can easily show that from the minimum space-time intervals, we
can deduce Planck's Law itself\cite{r32}. As the frequency is
inversely proportional to the time period, we can obtain a discrete
spectrum which is Planck's Law by the following derivation which is similar
to the usual theory\cite{r34}.\\
Let the energy be given by
$$E = g(\nu)$$
Then, $f$ the average energy associated with each mode is given by,
$$f = \frac{\sum_\nu g(\nu) e^{-g(\nu)/kT}}{\sum_\nu e^{-g(v)/kT}}$$
Again, as in the usual theory, a comparison with Wien's functional
relation, gives,
$$f = \nu F (\nu/kT),$$
whence,
$$E = g(\nu)\alpha \nu,$$
which is Planck's law.\\
Yet another way of looking at it is, as the momentum and frequency of the
classical oscillator have discrete spectra so does the energy.

\end{document}